\documentclass[twocolumn]{aastex61}
\usepackage{epsfig}
\usepackage{epstopdf}
\usepackage{amsmath}

\shorttitle{Wolf 503b}
\shortauthors{}

\definecolor{orange}{rgb}{0.8,0.4,0}
\newcommand\mysim{\mathord{\sim}}
\newcommand\kms{\ifmmode{\rm km\thinspace s^{-1}}\else km\thinspace s$^{-1}$\fi}
\newcommand\ms{\ifmmode{\rm m\thinspace s^{-1}}\else m\thinspace s$^{-1}$\fi}
\newcommand\vsini{\ifmmode{v\sin{i_\star}}\else $v\sin{i_\star}$\fi}
\newcommand\logg{\ifmmode{\log{g}}\else $\log{g}$\fi}
\newcommand\teff{\ifmmode{T_{\rm eff}}\else $T_{\rm eff}$\fi}
\newcommand{\feh}{\mbox{$\rm{[Fe/H]}$}}

\begin{document}

\title{A 2 Earth Radius Planet Orbiting The Bright Nearby K-Dwarf Wolf~503}

\correspondingauthor{Merrin S. Peterson}
\email{merrin.peterson@umontreal.ca}

\author{Merrin S. Peterson} 
\affil{D\'{e}partement de Physique, and Institute for Research on Exoplanets, Universit\'{e} de Montr\'{e}al, Montreal, H3T J4, Canada}

\author{Bj\"{o}rn Benneke} 
\affil{D\'{e}partement de Physique, and Institute for Research on Exoplanets, Universit\'{e} de Montr\'{e}al, Montreal, H3T J4, Canada}

\author{Trevor J.\ David}
\affil{Jet Propulsion Laboratory, California Institute of Technology, 4800 Oak Grove Drive, Pasadena, CA 91109, USA}

\author{Courtney D.\ Dressing}
\affil{Department of Astronomy, University of California, Berkeley, CA 94720}

\author{David Ciardi}
\affil{Caltech/IPAC-NASA Exoplanet Science Institute, 770 S. Wilson Ave, Pasadena, CA 91106, USA}

\author{Ian J.~M.\ Crossfield}
\affil{Department of Physics, and Kavli Institute for Astrophysics and Space Research, Massachusetts Institute of Technology, Cambridge, MA 02139, USA}

\author{Joshua E.\ Schlieder}
\affil{NASA Goddard Space Flight Center, 8800 Greenbelt Road, Greenbelt, MD 20771, USA}

\author{Erik A.\ Petigura}
\altaffiliation{NASA Hubble Fellow}
\affil{Department of Astronomy, California Institute of Technology, Pasadena, CA 91125, USA}

\author{Eric E.\ Mamajek}
\affil{Jet Propulsion Laboratory, California Institute of Technology, 4800 Oak Grove Drive, Pasadena, CA 91109, USA}

\author{Jessie L.\ Christiansen}
\affil{Caltech/IPAC-NASA Exoplanet Science Institute, 770 S. Wilson Ave, Pasadena, CA 91106, USA}

\author{Sam Quinn}
\affil{Harvard-Smithsonian Center for Astrophysics, 60 Garden Street, Cambridge, MA 02138, USA}

\author{Benjamin J.\ Fulton}
\affil{Caltech/IPAC-NASA Exoplanet Science Institute, 770 S. Wilson Ave, Pasadena, CA 91106, USA}

\author{Andrew W.\ Howard}
\affil{Department of Astronomy, California Institute of Technology, Pasadena, CA 91125, USA}

\author{Evan\ Sinukoff}
\affil{Department of Astronomy, California Institute of Technology, Pasadena, CA 91125, USA}

\author{Charles Beichman}
\affil{Caltech/IPAC-NASA Exoplanet Science Institute, 770 S. Wilson Ave, Pasadena, CA 91106, USA}

\author{David W.\ Latham}
\affil{Center for Astrophysics, 60 Garden Street, Cambridge, MA 02138, USA}

\author{Liang Yu}
\affil{Department of Physics, and Kavli Institute for Astrophysics and Space Research, Massachusetts Institute of Technology, Cambridge, MA 02139, USA}

\author{Nicole Arango}
\affil{College of the Canyons, Valencia, CA 91355, USA}

\author{Avi Shporer}
\affil{Department of Physics, and Kavli Institute for Astrophysics and Space Research, Massachusetts Institute of Technology, Cambridge, MA 02139, USA}

\author{Thomas Henning}
\affil{Max Planck Institute for Astronomy, 69117 Heidelberg, Germany}

\author{Chelsea X.Huang}
\affil{Department of Physics, and Kavli Institute for Astrophysics and Space Research, Massachusetts Institute of Technology, Cambridge, MA 02139, USA}

\author{Molly R.\ Kosiarek}
\altaffiliation{NSF Graduate Student Fellow}
\affil{Department of Astronomy and Astrophysics, University of California, Santa Cruz, CA 95064, USA}

\author{Jason Dittmann}
\affil{Department of Physics, and Kavli Institute for Astrophysics and Space Research, Massachusetts Institute of Technology, Cambridge, MA 02139, USA}

\author{Howard Isaacson}
\affil{Department of Astronomy, University of California, Berkeley, CA 94720}

\begin{abstract}

Since its launch in 2009, the \textit{Kepler} telescope has found thousands of planets with radii between that of Earth and Neptune. Recent studies of the distribution of these planets have revealed a rift in the population near 1.5--2.0\,$R_\Earth$, informally dividing these planets into ``super-Earths" and ``sub-Neptunes". The origin of this division is not well understood, largely because the majority of planets found by Kepler orbit distant, dim stars and are not amenable to radial velocity follow-up or transit spectroscopy, making bulk density and atmospheric measurements difficult. Here, we present the discovery and validation of a newly found $2.03^{+0.08}_{-0.07}~R_{\Earth}$ planet in direct proximity to the radius gap, orbiting the bright ($J=8.32$~mag), nearby ($D=44.5$~pc) high proper motion star Wolf~503 (EPIC 212779563). We classify Wolf~503 as a K3.5V star and member of the thick disc population. We determine the possibility of a companion star and false positive detection to be extremely low using both archival images and high-contrast adaptive optics images from the Palomar observatory. The brightness of the host star makes Wolf~503b a prime target for prompt radial velocity follow-up, \textit{HST} transit spectroscopy, as well as detailed atmospheric characterization with \textit{JWST}. With its measured radius near the gap in the planet radius and occurrence rate distribution, Wolf~503b offers a key opportunity to better understand the origin of this radius gap as well as the nature of the intriguing populations of ``super-Earths" and ``sub-Neptunes" as a whole.
\end{abstract}
    
    


\section{Introduction} \label{sec:intro}

The majority of close-in planets found by NASA's \textit{Kepler} satellite throughout the past decade are smaller than Neptune, but larger than Earth \citep{Batalha2013} \citep{Mullally15} \citep{Howard13}. The \textit{Kepler} and \textit{K2} missions have shown us that, of the planets within our detection limits ($P>100$ days, $R_p>1.0R_{\Earth}$), these smaller planets are by far the most common in the galaxy \citep{Fressin13} \citep{Fulton2017}, though there is no analog in the solar system from which this could have been predicted.

A drop in the population of planets at radii larger than $4.0~R_\Earth$ (i.e., larger than Neptune) is satisfactorily explained by runaway gas accretion \citep{Bate2003} \citep{Mordasini09}. Larger planets are massive enough to accrete H and He from the protoplanetary disc, becoming puffy and increasing in radius. However, refined studies of the distribution of planets within the $1-4~R_\Earth$ range have revealed a significant drop in the population, or ``Fulton gap" between $1.5-2.0~R_\Earth$ \citep{Fulton2017} \citep{Owen13}, which is not yet well-understood. 

\begin{figure*}[ht]
\begin{center}
\includegraphics[width=2.0\columnwidth]{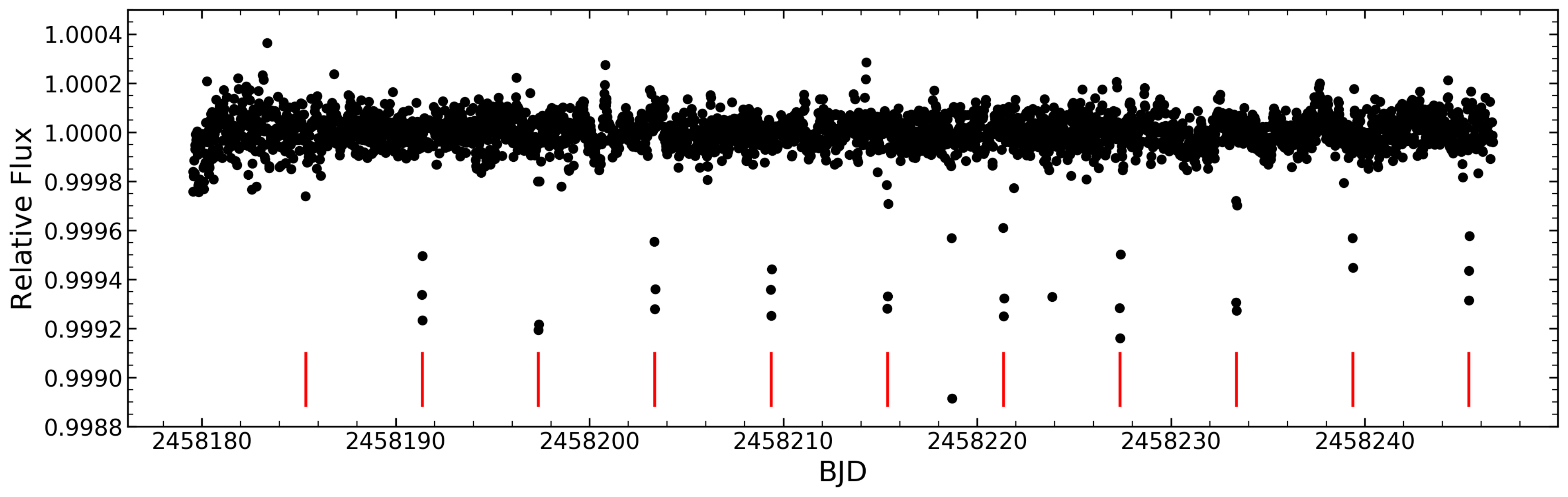}
\caption{Extracted light curve for Wolf~503 (EPIC 212779563). Transit times according to our fit are indicated with a red line. The first observed transit is not easily visible in this plot because the transit coincided with a thruster burn during which two data points were flagged and removed (see Fig.~\ref{fig:individual_fits}). \label{fig:full_curve}}
\end{center}
\end{figure*}

Photoevaporation presents a possible explanation for the gap, and is a particularly important factor for the close-in planets preferentially detected by \textit{Kepler}. Planets with radii between $1.5$ and $2.0~R_\Earth$ could represent a relatively rare group of planets retaining thin atmospheres, while super-Earths are photoevaporated rocky bodies and the sub-Neptunes are massive enough to retain thick atmospheres \citep{Lopez16}. 
It has also been postulated that the sub-Neptunes form earlier in the evolution of the protoplanetary disc than super-Earths, when there is still more gas in the protoplanetary disc, giving them thicker atmospheres and larger radii \citep{Lee2014}. The gap would then represent an intermediate stage in disc evolution in which planets are not likely to form. 


Explanations for the bimodal distribution of planets which invoke composition should be tested with mass (i.e., bulk density) measurements and transit spectroscopy to determine the composition and atmospheric mass fraction of planets on both sides of the rift.

However, planets favorable for these detailed follow-up characterizations are missing. Although \textit{Kepler} has found thousands of bona fide $1-4~R_\Earth$ planets, due to the satellite's due to Kepler’s ~100 sq. deg. field-of-view, relatively few bright stars were targeted and most Kepler planet hosts are distant and dim. As is shown in Fig.~\ref{fig:Hmag_vs_Rp}, Wolf~503b joins only a handful of bright targets at its size. With so few photons, the detailed spectra required to make quality mass and atmospheric composition measurements are often impossible to obtain, and although there has been much effort to constrain the density of planets in this region \citep{Dumusque14} \citep{Weiss2014} \citep{Rogers15}, the parameter space near the Fulton gap remains relatively unexplored.

In this work, we present the detection and validation of a newly found $2.0~R_\Earth$ planet from \textit{K2} which represents one of the best opportunities to date to conduct a detailed radial velocity and atmospheric study of a planet in the 1-4$R_{\Earth}$ range. In Sec.~\ref{sec:photometry} we describe the collection and calibration of the \textit{K2} photometry, as well as our detection pipeline. In Sec.~\ref{sec:stellarHistory} we discuss the research history of the host star and its galactic origins. We obtain our own spectrum of Wolf~503, classify the star and determine stellar parameters in Sec.~\ref{sec:spectroscopy}. Our methods of target validation are described in Sec.~\ref{sec:validation} and the final light curve fitting and results are found in Sec.~\ref{sec:fitting} 

\begin{table}
\caption{Stellar Parameters}
    \centering
    \footnotesize
    \begin{tabular}{ m{2.4cm} m{2.2cm} m{2.4cm} } 
         \hline
         \hline
         Parameter & Value & Source \\
         \hline
         \multicolumn{2}{c}{Identifying Information} & \\ [0.1ex]
         \hline
         $\alpha$ R.A. (hh:mm:ss) J2000 & 13:47:23.4439 & \\
         $\delta$ Dec. (dd:mm:ss) J2000 & -06:08:12.731 & \\
         EPIC ID & 212779563 & \\
         \multicolumn{2}{c}{Photometric Properties} & \\
         \hline
         B (mag)..... & $11.30\pm0.01$ & \citep{Mermilliod87}\\
         V (mag)..... & $10.28\pm0.01$ & \citep{Mermilliod87}\\
         G (mag)..... & $9.808\pm0.001$ & Gaia DR1\\
         J (mag)..... & $8.324\pm0.019$ & 2MASS\\
         H (mag)..... & $7.774\pm0.051$ & 2MASS\\
         K (mag)..... & $7.617\pm0.023$ & 2MASS\\
         \multicolumn{2}{c}{Spectroscopic and Derived Properties} & \\
         \hline
         $\mu_{\alpha}$ (mas yr$^{-1}$) & $-343.833\pm0.073$ & Gaia DR2\\
         $\mu_{\delta}$ (mas yr$^{-1}$) &  $-573.134\pm0.073$ & Gaia DR2\\
         Barycentric rv \newline (km s$^-1$) & $-46.826\pm0.015$ & Gaia DR2\\
         Distance (pc) & $44.583\pm0.096$ & Gaia DR2\\
         Age (Gyr) & $11\pm2$ & This Paper \\
         Spectral Type & $K3.5\pm0.5$V & This Paper\\
         $[Fe/H]$   & $-0.47\pm0.08$ & This Paper\\
         $logg$ (K) & $4.62^{+0.02}_{-0.01}$ & This Paper\\
         $\teff$ (K) & $4716 \pm 60$ & This Paper\\
         $M_*$ ($M_{\odot}$) & $0.688^{+0.023}_{-0.016}$ & This Paper\\
         $R_*$ ($R_{\odot}$) & $0.690^{+0.025}_{-0.024}$ & This Paper\\
         $L_*$ ($L_{\odot}$) & $0.227^{+0.009}_{-0.010}$ & This Paper\\
         \hline
    \end{tabular}
    \label{tab:stellar}
\end{table}

\section{Observations and Analysis} \label{sec:obs}
Identified as a planet candidate from C17 of \textit{K2}, Wolf~503 was recognized as an excellent host for follow-up study, being both bright ($K_p=9.9$) and nearby (45pc). Here we present the treatment of the photometry used to detect Wolf~503b, as well as our planet validation techniques, and derive both planetary and stellar parameters.

\begin{figure*}[ht]
\begin{center}
\includegraphics[width=2.0\columnwidth]{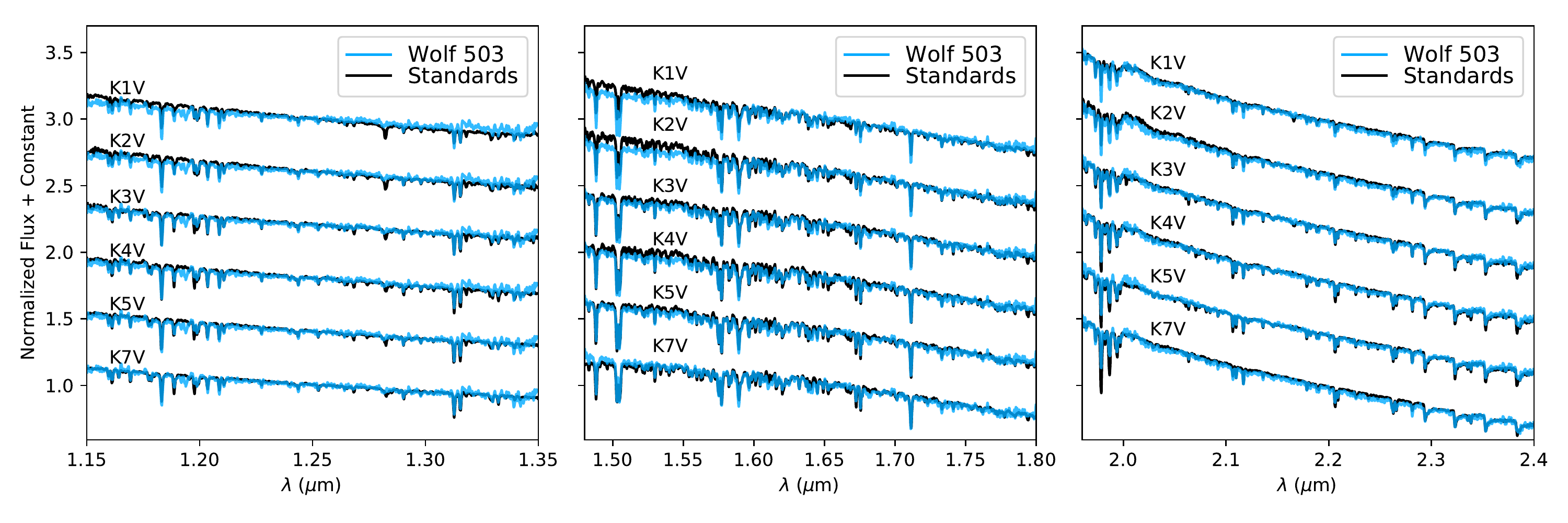}
\caption{Final, calibrated SpeX spectra for Wolf~503 shown compared to spectral standards. We find the best visual match for Wolf~503 indicates a K3$\pm1$V spectral type, consistent with previous classifications \citep{Pickles10}. \label{fig:spectrum}}
\end{center}
\end{figure*}

\subsection{Photometry Extraction and Transit Detection} \label{sec:photometry}

The photometric extraction and transit detection methods used to identify Wolf~503b are the same as those applied to all light curves in C17 and are described in our corresponding C17 summary paper Crossfield et al. 2018 (submitted). As \textit{K2} operates using only two of \textit{Kepler}'s four initial reaction wheels, the telescope drifts along its roll axis by a few pixels every several days, and thruster fires are used to maintain the telescope's pointing. The change in flux resulting from this drift is removed by fitting the flux as a function of position along the drift path, which is highly similar between thruster fires. However, data acquired during these thruster burns is not reliable and is masked out, as in the first transit of the light curve for Wolf~503, shown in Fig.~\ref{fig:full_curve}. 

With the extracted light curve, we detected a candidate at $P=6.0$ days with $S/N=38$ having 11 transits throughout the time of observation. The candidate was marked as a particularly intriguing KOI for the properties of its host star following the manual vetting procedure of the C17 candidates.



\begin{figure}
\begin{center}
\includegraphics[width=1.0\columnwidth]{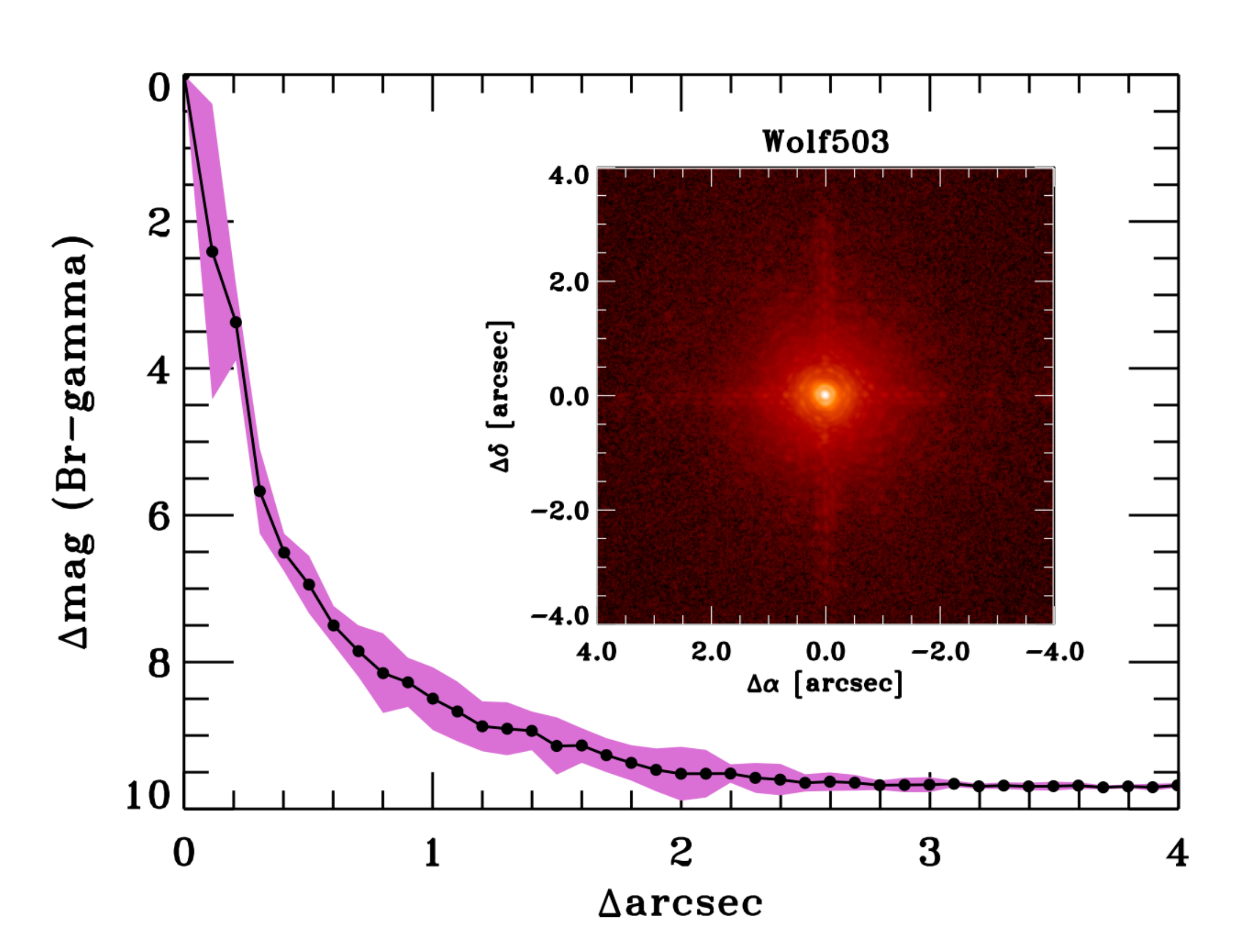}
\caption{Contrast sensitivity and inset image of Wolf~503 in Br-$\gamma$ as observed with the Palomar Observatory Hale Telescope adaptive optics system, The $5\sigma$ contrast limit is plotted against angular separation in arcseconds (fill circles).  The shaded region represents the dispersion in the sensitivity caused by the azimuthal structure in the image (inset).\label{fig:AO}}
\end{center}
\end{figure}

\begin{figure*}[ht]
\begin{center}
\includegraphics[width=2.0\columnwidth]{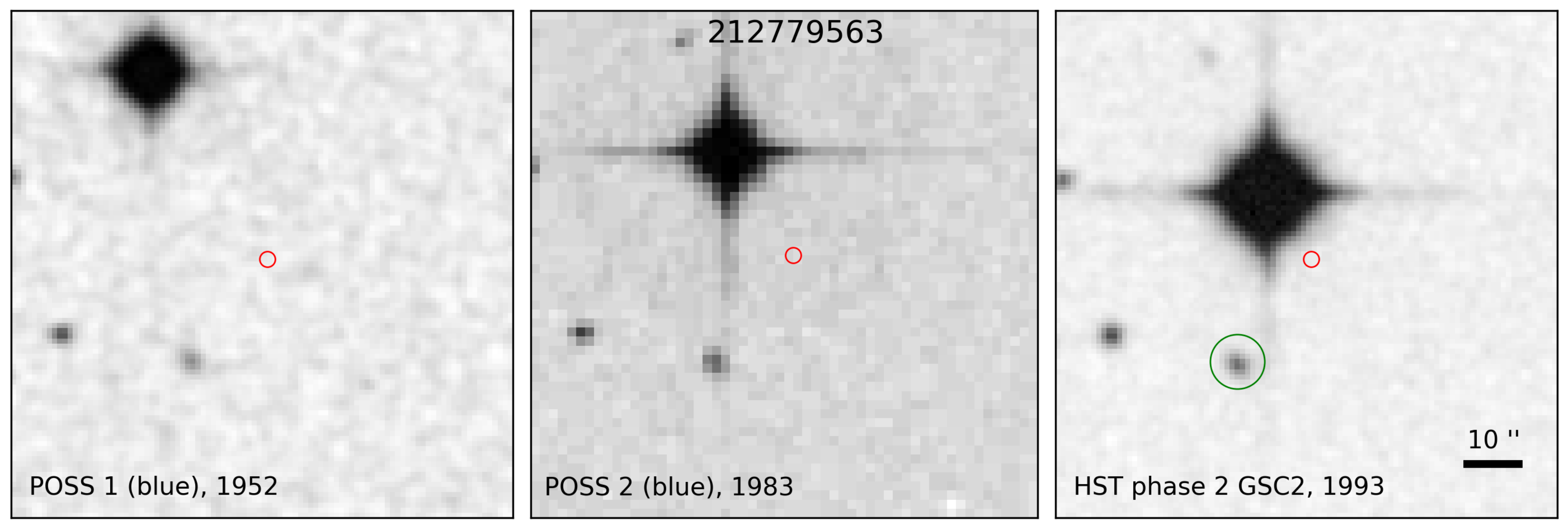}
\caption{Archival images from the blue plate of the POSS I (May 23, 1952), II (May 7, 1983) and Guide Star Catalog II sky surveys (taken on HST May 7, 1983). Wolf~503's significant high proper motion is clear in the sequence of images, and there are no background sources detected at its 2018 location, marked in red. The nearest source is the faint galaxy LCRS B134447.1-055347, circled in green in the right panel, which is both 10 magnitudes fainter than Wolf~503 and found outside our extraction aperture.
\label{fig:archival_images}}
\end{center}
\end{figure*}

\subsection{Activity, Age and Membership} \label{sec:stellarHistory}
Wolf~503 (BD-05 3763, MCC 147, LHS 2799, G 64-24, HIP 67285, TYC
4973-1501-1, 2MASS J13472346-0608121) has been a sparsely studied
nearby cool star since its discovery a century ago as a high
proper motion star by
\citet{Wolf1919}.
The star subsequently appeared in several catalogues of high proper
motion over the past century, Ci 20 806 \citep{Porter30}, G 64-24
\citep{Giclas63}, with Wilhelm Luyten designating the star no fewer
than six times in his proper motion catalogs \footnote{ entry \#402 in
\citet{Luyten23} (stars with motions exceeding 0\arcsec.5/yr), \ LPM
492 \citep{Luyten41}, LFT 1037 \citep{Luyten55}, LHS 2799
\citep{Luyten79}, and as NLTT 35228 and LTT 5351 in \citet{Luyten80}}.

The star was classified in numerous spectral surveys, 
 as a K5V by \citet[][; identified as UPG 336]{Upgren72}, and
\citet{Bidelman85} published Kuiper's posthumous classification for
the star as K4 from his 1937-1944 survey.  Of the K5V stars from
Upgren et al.'s (1972) survey, 89 were later classified in the CCD
optical spectroscopy surveys of nearby stars by \citet{Gray03} and
\citet{Gray06}, and assigned an average type of K4.6 ($\pm$0.1 subtype
s.e.m., $\pm$1.1 subtype rms).  \citet{Pickles10} found that the best
fit template for the $B_T V_T J H K_s$ photometry was that for a K4V
star. The combined previously published values suggest a spectral type of K4V.


Recently Gaia DR2 has provided an
ultra-precise trigonometric parallax ($\varpi$ =
22.430\,$\pm$\,0.048 mas; corresponding to $d$ = 44.583\,$\pm$\,0.096 pc), as well as precise proper motion and radial velocity measurements, which are listed in Table \ref{tab:stellar}. Gaia itself measured a radial velocity of
-46.64\,$\pm$\,0.50 km\,s$^{-1}$ (2 observations), and independently, \citet{Sperauskas16} reported radial velocity of -47.4\,$\pm$\,0.7km\,s$^{-1}$ based on 2 CORAVEL measurements over 98 days.  
Combining
the Gaia DR2 position, proper motion, and parallax, and the mean Gaia
DR2 ground-based radial velocity (from HARPS), we estimate barycentric
space velocity of $U, V, W$ = -25.21, -116.86, -88.44 ($\pm$0.18,
0.21, 0.13) km\,s$^{-1}$ (total velocity 148.71\,$\pm$\,0.18
km\,s$^{-1}$), where $U$ is towards the Galactic center, $V$ is in the
direction of Galactic rotation, and $W$ is towards the north Galactic
pole \citep{ESA97}.  Using the velocity moments and local stellar
population densities from \citet{Bensby03}, this $UVW$ velocity is consistent with the following membership probabilities:
$<$10$^{-5}$\%, 81\%, 19\%, for the thin disk, thick disk, and halo,
respectively, highly indicative of  membership to the thick disc population. 

\citet{Mikolaitis17} analyzed high resolution high S/N
HARPS spectra and found the star to be fairly metal poor ([Fe/H]
$\simeq$ -0.37 based on two pairs of [Fe I/H] and [Fe II/H]
abundances).  Its combination of low metallicity, supersolar [Mg I/Fe]
($\sim$0.28) and [Zn I/Fe] (0.19), and subsolar [Mn I/Fe]
($\sim$-0.16), led \citet{Mikolaitis17} to chemically classify the
star as belonging to the thick disk.  Hence there is both kinematic
and chemical abundance data for Wolf~503 consistent with its
membership to the thick disk. The thick disk shows a metallicity-age
gradient \citep[e.g.][]{Bensby04}, and given Wolf~503's combination of
[Fe/H] and [Mg/Fe] compared to age-dated thick disk members
\citep{Haywood13}, it is likely in the age range $\sim$9-13 Gyr. Hence
we adopt 11\,$\pm$\,2 Gyr for Wolf~503.

\subsection{Spectroscopy and Stellar Parameters}\label{sec:spectroscopy}

We obtained an $R\approx2000$ infrared spectrum of Wolf~503 covering the spectra range between $0.7-2.55 \micron$ at the NASA Infrared Telescope Facility (IRTF). We use the SpeX spectrograph in SXD mode with the 0.3" x 15" slit. The spectrum was taken June 3, 2018, on a partly cloudy night with an average seeing of $0.6\arcsec$. Reduction of the spectrum was performed with the SpeXTool \citep{Cushing2005} and xtellcor \citep{Vacca03} software packages as in \citet[][]{Dressing17}. The sky subtraction was performed using a nearby A star, HD 122749, observed immediately after Wolf~503b, at a similar airmass. Before performing our spectral analysis, we corrected for the radial velocity of the target and barycentric velocity of IRTF. The final spectrum is shown in Fig.~\ref{fig:spectrum}. The best match indicates a spectral type of $K3.5\pm1$V suggesting an effective temperature of approximately $4750\pm100\,K$ for SpeX spectrum.

During the vetting of candidates from C17 of $K2$ described in Crossfield et al. 2018 (submitted), a spectrum was also obtained from the Tillinghast Reflector Echelle Spectrograph \citep[TRES;][]{furesz:2008} mounted on the 1.5-m Tillinghast Reflector at Fred Lawrence Whipple Observatory on Mount Hopkins was
obtained on UT 2018 May 23. TRES is a fiber-fed, cross-dispersed echelle spectrograph with a resolving power of $R \mysim 44,000$, a wavelength coverage of $3850$--$9100$\,\AA, and radial-velocity stability of $10$\ to $15$\,\ms. The spectrum was reduced and optimally extracted, and wavelength calibrated according to the procedure described in \citet{buchhave:2010}, and we derived stellar atmospheric parameters
using the Stellar Parameter Classification code \citep[SPC;][]{buchhave:2012}. We find $\teff = 4640 \pm 50$\,K, $\logg = 4.68 \pm 0.10$, $\feh = -0.47 \pm 0.08$, and $\vsini = 0.8 \pm 0.5$. We note that SPC determines the stellar parameters using synthetic spectra with a fixed macroturbulence of $1$\,\kms, which may bias \vsini\ measurements of slow rotators like this one. Regardless, Wolf 503 has a low projected rotational velocity, as is expected for an old K dwarf, which bolsters its status as a good candidate for
precise radial velocity observations. We derive an absolute radial velocity of $-46.629 \pm 0.075$\,\kms.

We conclude that the SpeX spectrum and the TRES spectrum result in consistent estimates of the stellar temperature. These values are also consistent with the value from the PASTEL catalogue of 4759 K \citep{soubiran_pastel_2010} as well as Wolf~503's colors ($B-V=1.02$, $V-K=2.66$), leading us to adopt the spectral type of $K3.5\pm0.5V$. 

Finally, we adopt $\teff=4716\pm60$~K, the average and scatter of the three spectroscopic values, as our final value for the stellar temperature. We then calculate the stellar parameters using Isoclassify \citep{huber_asteroseismology_2017}. We adopt the $\logg$ and $\feh$ from the TRES spectrum, as well as the K magnitude. We use the K magnitude because it is least affected by extinction and inflate the error bars of the K magnitude to account for the uncertainty extinction. We determine the best stellar radius estimate using the direct method in Isoclassify \citep{huber_asteroseismology_2017}. We obtain the stellar mass using the grid mode. The resulting stellar parameters are listed in Table \ref{tab:stellar}.

\subsection{Target Validation}\label{sec:validation}

By far the most pernicious false positives detected by \textit{K2} are eclipsing binaries, which may closely resemble exoplanet transits at grazing incidence, or when the binary system is found in the background of a brighter star \citep{Morton12}. We used archival and adaptive optics images to investigate the possibility of a false positive detection due to a companion star or background sources, and find no source in the vicinity of Wolf~503 which could have contaminated our detection.

\begin{figure*}[ht]
\begin{center}
\includegraphics[width=2.0\columnwidth]{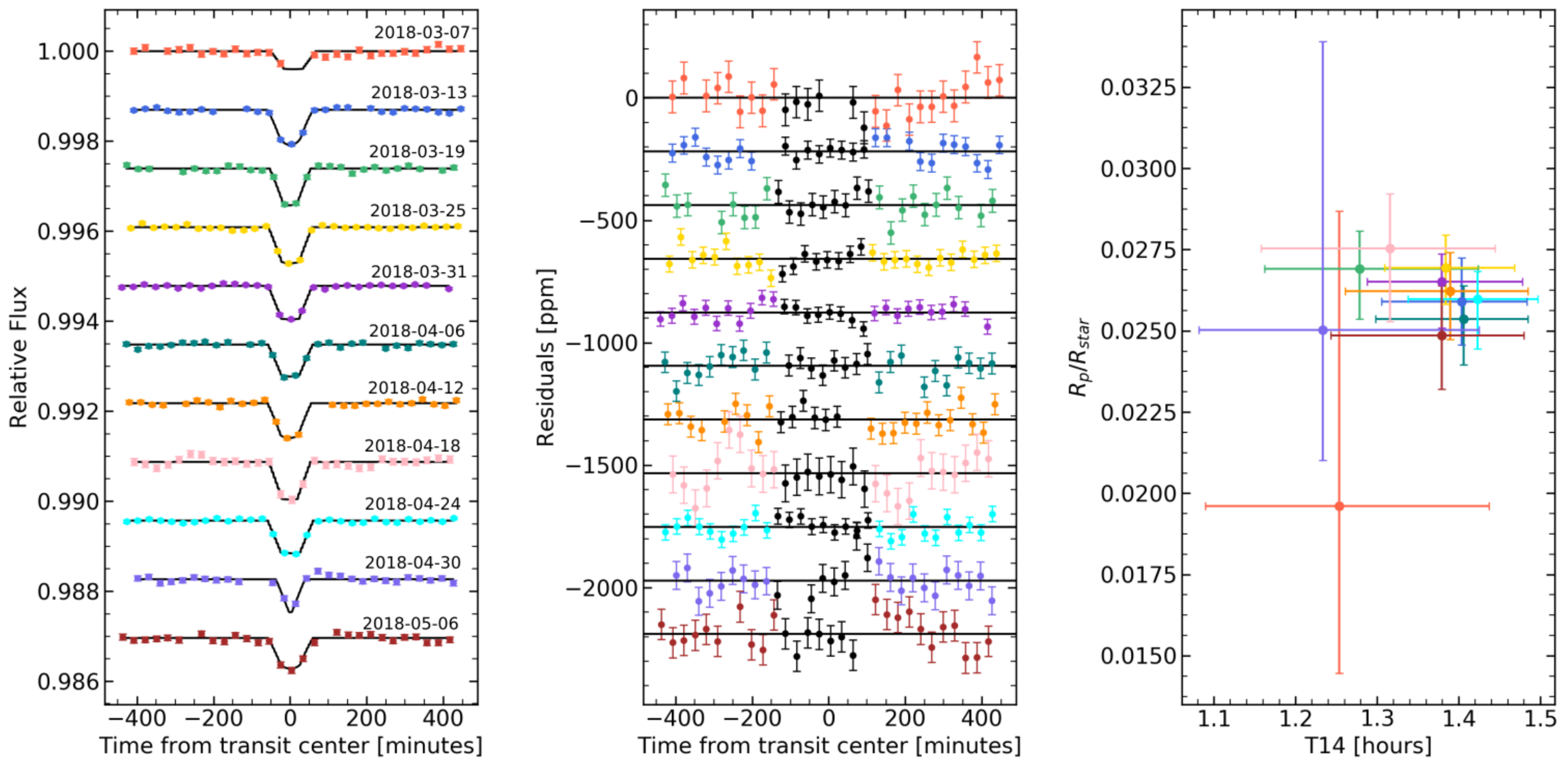}
\caption{Individual transit fits of Wolf~503b. The left panel shows each individual transit with its corresponding best fit model. The residuals are shown in the center panel, with the residuals in the range $T_0 \pm T_{14}$ marked in black. The right plot shows the best guess and 1$\sigma$, or 68\% confidence limits on the $R_p/R_*$ and $T_{14}$ parameters, which are consistent for all transits, further support that the signal is consistent with a transiting planet. Uncertainties on the first and tenth transits (red and violet) are higher due to masked data points coinciding with a thruster burn near the time of the transit. \label{fig:individual_fits}}
\end{center}
\end{figure*}

\subsubsection{Adaptive Optics}

Wolf~503 was observed on the night of UT 2018 June 01 UT at Palomar Observatory with the 200\arcsec\ Hale Telescope using the near-infrared adaptive optics (AO) system P3K and the infrared camera PHARO (Hayward et al. 2001). PHARO has a pixel scale of 0.025\arcsec\ per pixel with a full field of view of approximately 25\arcsec. The data were obtained with a narrow-band Br-$\gamma$ filter ($\lambda_o = 2.18;\ \Delta_\lambda = 0.03\ \micron$).

The AO data were obtained in a five-point quincunx dither pattern with each dither position separated by 4\arcsec. Each dither position is observed three times, each offset from the previous image by 0.5\arcsec\ for a total of 15 frames; the integration time per frame was 4.428 s for a total of 66 on-source integration time. We use the dithered images to remove sky background and dark current, and then align, flatfield, and stack the individual images. The final PHARO AO data have a FWHM of 0.099\arcsec.

The sensitivities of the final combined AO image were determined by injecting simulated sources azimuthally around Wolf~503 every 45$^\circ$ at separations of integer multiples of the central source.  The brightness of each  injected  source  was  scaled  until  standard  aperture photometry detected it with 5$\sigma$ significance. The resulting brightness of the injected sources relative to Wolf~503 set the contrast limits at that injection location. The average 5$\sigma$ limits and associated rms dispersion caused by azimuthal asymmetries from residual speckles as a function of distance from the primary target are shown in Fig.~\ref{fig:AO}.

The AO imaging revealed no additional stars within the limit of 0.099\arcsec. For a binary system at a distance of $44.58$~pc, this limits the separation of a possible binary to less than 4.4 AU. According to the distribution of binary star systems found in \citet{Raghavan10}, only 12\% of stars are found in such systems. Furthermore, we find the light curve properties (discussed in Sec.~\ref{sec:discussion}) inconsistent with an eclipsing binary, except in the case of a multiple star system featuring two smaller companions in a 6.0 day orbital period, in which one companion star were completely eclipsed by the other. We consider such a unique multiple-star system far less likely than a single transiting planet. Additionally, most $0.1M_{\odot}$ eclipsing binary companions orbiting within 4.4 AU would induce a radial velocity amplitude on the order of 15 km/s, of which there is no indication through years of radial velocity measurements. We determine the likelihood of a false positive due to a bound companion to be extremely low.

\subsubsection{Archival Images}

Even in the absence of a nearby contaminant, adaptive optics cannot eliminate the possibility of a background source directly behind the target, which could be responsible for the signal itself, or would otherwise decrease the apparent transit depth. To address this, we exploit archival imaging from the Palomar Observatory Sky Surveys I, II and Guide Star Catalogue 2 surveys. Fig.~\ref{fig:archival_images} shows the present-day location of Wolf~503 in each of the 3 surveys. The blue plate from POSS I (taken May 23, 1952) and the HST image from GSC2 (taken March 29, 1993 with HST) have a 1\arcsec pixel scale, and the blue plate from POSS II (taken May 7, 1983) has a 0\arcsec.59 pixel scale.

The nearest object detected to Wolf~503's 2018 location is the galaxy LCRS B134447.1-055347, which is located $\approx25.1 ''$ from the target, placing it outside the aperture used in our extraction. Moreover, the galaxy has a Gaia magnitude of 19.6: being both 10 magnitudes fainter and outside the aperture, we find no background sources which may influence our photometry.

\begin{table}
\caption{Planet Parameters}
    \footnotesize
    \centering
    \begin{tabular}{c c c}
         \hline \hline
         Parameter & Units & Value\\
         \hline
         $T_0$ & BJD$_{TBD}$ - 2457000 & $1185.36087\substack{+0.00053 \\ -0.00038}$\\ 
         $P$ & day & $6.00118\substack{+0.00008 \\ -0.00011}$\\
         $R_p/R_{*}$ & \% & $2.694\substack{+0.026 \\ -0.026}$\\
         $T_{14}$ & hr & $1.321\substack{+0.051 \\ -0.039}$\\
         $b$ & - & $0.387\substack{+0.067 \\ -0.061}$\\
         $R_p$ & $R_{\Earth}$ & $2.030^{+0.076}_{-0.073}$\\
         $a$ & AU & $0.0571\pm0.0020$\\
         $S$ & $S_\Earth$ & $69.6\pm3$\\
         $T_{\mathrm{eq,A=0}}$ & K & $805\pm9$\\
         \hline
    \end{tabular}
    \label{tab:planet}
\end{table}

\begin{figure}
\begin{center}
\includegraphics[width=1.0\columnwidth]{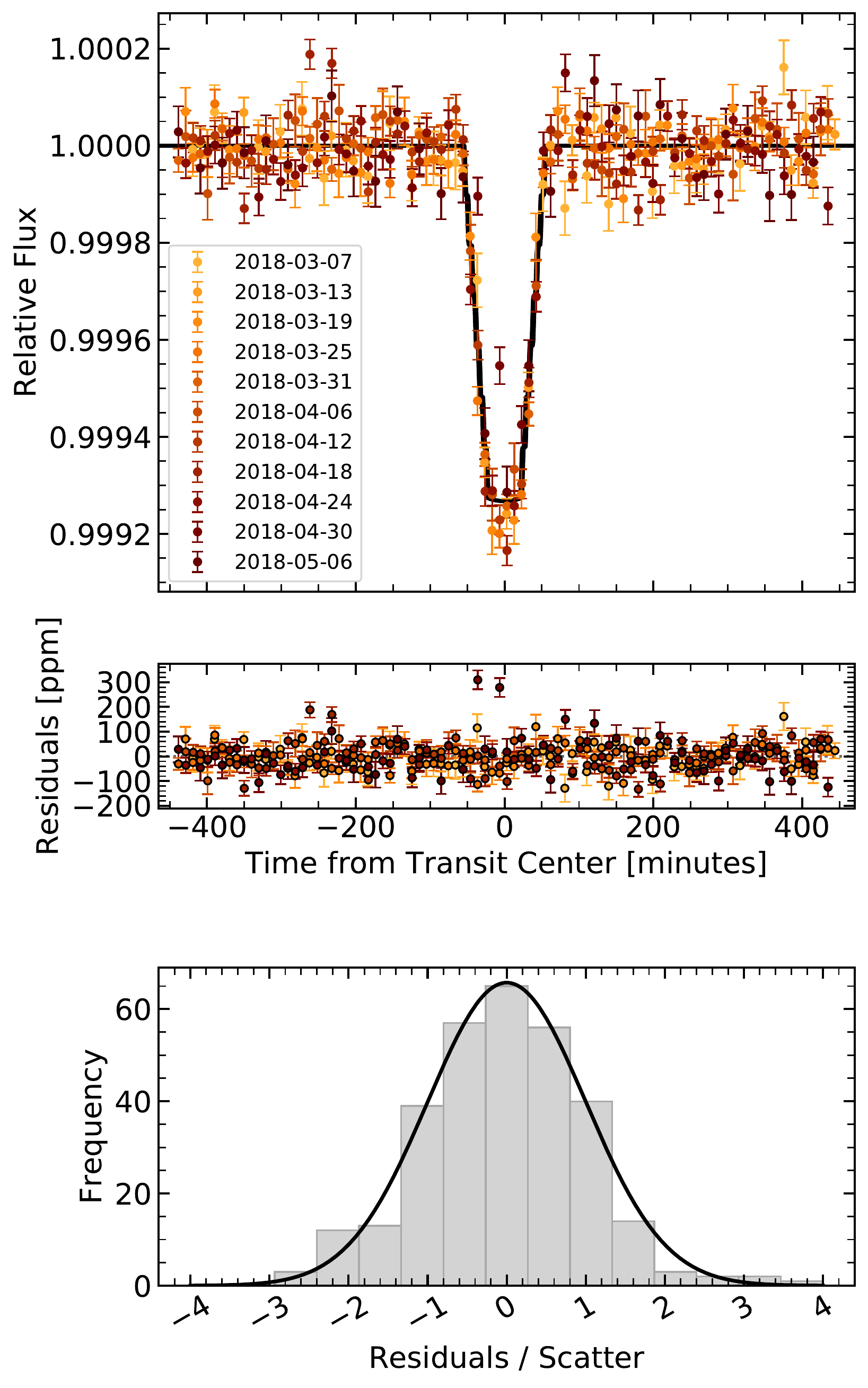}
\caption{Final light curve fit from ExoFit for the combined 11 transits. In the top panel, the best fit is shown in black with the detrended light curves for each transit. Accounting for the 30 minute cadence of the \textit{K2} data gives the best fit its trapezoidal shape. The residuals are plotted in the middle panel, and are binned in the bottom panel histogram by the number of $\sigma$ from the best fit, where they follow a standard normal distribution of the same area. \label{fig:pretty_plot}}
\end{center}
\end{figure}

\subsection{Light Curve Fitting} \label{sec:fitting}

We fit the light curve of Wolf~503 using ExoFit, a modular light curve analysis tool developed for the joint analysis of data from \textit{Kepler}, \textit{Spitzer}, and \textit{HST}. ExoFit jointly or individually fits transits and explores the parameter space using the Affine Invariant Markov Chain Monte Carlo (AI-MCMC) Ensemble sampler available through the emcee package in Python. Details can be found in \citet{Benneke2017}. 

We performed individual transit fits in addition to fitting the transits simultaneously. For all fits, we initialize the MCMC chains with uniform priors using the best fit values from TERRA, and fit the transit start time $T_0$, duration $T_{14}$, depth $R_p/R_{*}$, impact parameter $b$, limb darkening coefficient, as well as a linear background for each transit and scatter term. For the joint fit, we also fit the period $P$. In each fit, we assign 6 walkers for each parameter and find good convergence after 3000 steps, taking the initial 60\% as burn-in. 

The transits were first fit individually, and the resulting fits are shown in Fig. \ref{fig:individual_fits}. Of the 11 transits observed, all are consistent in $R_p/R_{*}$ and $T_{14}$. We obtain our best fitting planet parameters from a joint fit of the 11 transits using the initialization as previously described. 
The parameters resulting from this fit are summarized in Table \ref{tab:planet}, where the error in $R_p$ and $a$ are dominated by the stellar parameters, to which we have assigned conservative error estimates. The best fit light curve is shown in Fig.~\ref{fig:pretty_plot}, where the combined residuals are well-behaved. The regularity in depth and duration shown in Fig.~\ref{fig:individual_fits} is most consistent with a transiting planet, and we detect no even-odd variation indicative of an eclipsing binary. Furthermore, the best-fit in Fig.~\ref{fig:pretty_plot} is distinctly flat-bottomed, inconsistent with the V-shaped light curves characteristic of eclipsing binaries. This diluted, flat-bottomed shape could be reproduced by two smaller companions orbiting each other with a 6 day period. However, as discussed in Sec.~\ref{sec:validation}, in addition to being less likely than a single transiting planet, such a companion would induce a significant radial velocity which has not been detected. 

\section{Discussion} \label{sec:discussion}

\begin{figure}
\begin{center}
\includegraphics[width=1.0\columnwidth]{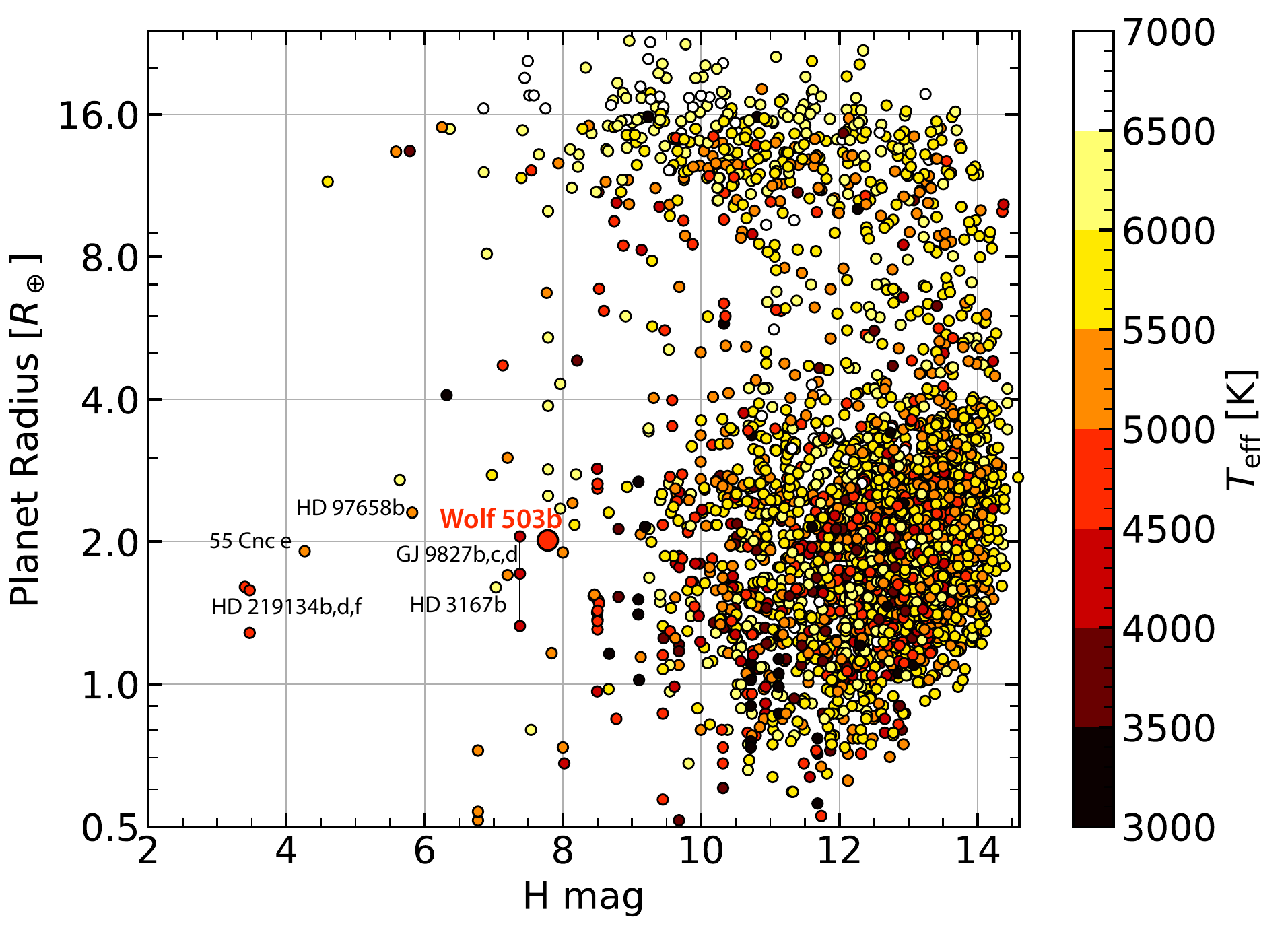}
\caption{Planet radius and stellar host magnitude of Wolf~503b (larger circle) in comparison to all planets at the NASA Exoplanet Archive (colored points). The color of the points indicates the stellar temperature. Planets in a similar size range orbiting bright stars are labeled. Wolf~503 is among the brightest systems with a planet near 2 $R_\Earth$ detected to date.\label{fig:Hmag_vs_Rp}}
\end{center}
\end{figure}

\begin{figure}
\begin{center}
\includegraphics[width=1.0\columnwidth]{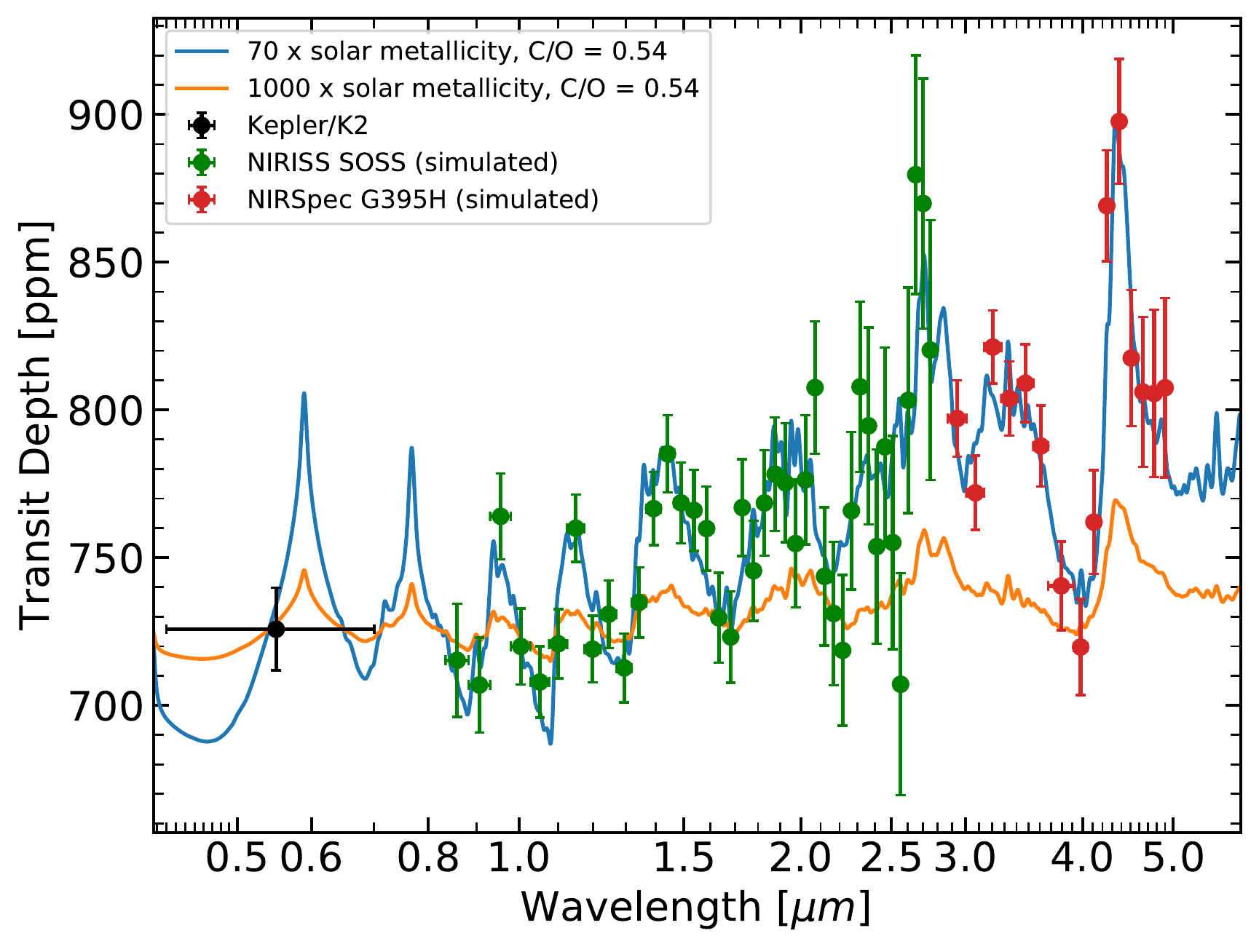}
\caption{Model transit spectra and simulated \textit{JWST} observations for Wolf~503b. Observations of a single transit with JWST/NIRISS (green) or JWST/NIRSpec (red) could readily detect molecular absorption for hydrogen-dominated, cloud free atmospheres (blue). The planetary mass assumed in the models is 5.3\,$M_{\Earth}$. Models are computed as described in \citet{benneke_atmospheric_2012} and \citet{benneke_strict_2015}. Simulated observational uncertainties are from PandExo \citep{batalha_pandexo_2017}.\label{fig:sim_spec}}
\end{center}
\end{figure}

From our combined imaging, photometric and spectral analyses, we establish Wolf~503b as a $2.03^{+0.08}_{-0.07}~R_{\Earth}$ planet orbiting its host star with a period of 6.0012 days. Wolf~503b is truly distinguished as its radius places it directly at the edge of the radius gap near 1.5--2.0\,$R_\Earth$, while its bright host star (H=7.77 mag, V=10.28 mag) makes it one of the best targets for radial velocity follow-up and transit spectroscopy at its size (Fig.~\ref{fig:Hmag_vs_Rp}).

Radial velocity measurements of Wolf~503b present an excellent opportunity to probe the bulk density of a planet just outside the radius gap. The amplitude of the expected RV signal depends strongly on the planet composition and amount of gas accreted. As Wolf~503b is similar in size to 55 Cnc e, though at a lower temperature, we investigate its composition using the mass-radius relationships for rocky compositions found in \citet{valencia_composition_2010} and \citet{Gillon12}. For the gas-poor scenario, the minimum mass required for a rocky composition (with no iron), is roughly $10\,M_{\Earth}$, with an Earth-like composition corresponding so $14\,M_{\Earth}$. These masses would result in RV amplitudes of roughly 4.5 and 6.3 m/s, higher than the RV amplitudes resulting from the gas-rich scenario. For a volatile planet with a 0.01\% H/He envelope, we would expect a mass of roughly $8\,M_{\Earth}$, whereas a 20\% water envelope would suggest $6\,M_{\Earth}$, and the empirical mass-radius relation by \citet{weiss_mass_2013} would suggest $5.3\,M_{\Earth}$, giving RV amplitudes of 3.6, 2.7, and 2.4 m/s. These amplitudes are detectable with existing precision radial velocity spectrographs, particularly for a bright target such as Wolf~503, and will provide critical constraints on the bulk composition of the planet.

Wolf~503b is also an ideal target for detailed characterization with \textit{JWST}. With $J=8.32$~mag, it is just below the saturation levels of $J>7$~mag and $J>6$~mag on the \textit{NIRISS} and \textit{NIRSpec} grisms. If Wolf~503b indeed harbours a thick atmosphere, it is one of the best known targets to date for transmission spectroscopy at its size. Fig.~\ref{fig:sim_spec} shows two simulated transit spectra for Wolf~503b, the blue corresponding to a hydrogen-rich, Neptune-like atmosphere and the orange corresponding to an atmosphere rich in water. Simulated \textit{NIRISS} and \textit{NIRSpec} data for the Neptune-like atmosphere is overplotted, demonstrating the high-confidence with which we will be able to constrain the structure and abundances of atmospheric molecules on Wolf~503b.

Both radial velocity measurements and atmospheric characterization with \textit{HST} would be valuable short-term follow-up to this work. Wolf 503b is among only a handful of planets in its size range for which this follow-up can be done efficiently today. As such, we expect Wolf 503b to play a critical role in providing near-term insights into distribution of core masses, envelope fraction, and the role of photoevaporation for planets near the Fulton gap. It can also serve as archetype to this class of small planets orbiting nearby stars in preparation for future characterization of similarly bright \textit{TESS} systems.

\acknowledgments

\bibliography{main.bbl}

\end{document}